# Scintillation correction for astronomical photometry on large and extremely large telescopes with tomographic atmospheric reconstruction

## J. Osborn[*]

*Department of Physics, Centre for Advanced Instrumentation, University of Durham, South Road, Durham DH1 3LE, UK*

### ABSTRACT

We describe a new concept to correct for scintillation noise on high-precision photometry in large and extremely large telescopes using telemetry data from adaptive optics (AO) systems. Most wide-field AO systems designed for the current era of very large telescopes and the next generation of extremely large telescopes require several guide stars to probe the turbulent atmosphere in the volume above the telescope. These data can be used to tomographically reconstruct the atmospheric turbulence profile and phase aberrations of the wavefront in order to assist wide-field AO correction. If the wavefront aberrations and altitude of the atmospheric turbulent layers are known from this tomographic model, then the effect of the scintillation can be calculated numerically and used to normalize the photometric light curve. We show through detailed Monte Carlo simulation that for an 8 m telescope with a 16 × 16 AO system we can reduce the scintillation noise by an order of magnitude.

## 1 INTRODUCTION

High-precision photometry is key to several branches of astrophysics, including (but not limited to) the study of extrasolar planets, stellar seismology and the detection of small Kuiper belt objects within our Solar system. The difficulty with such observations is that although the targets are bright, the variation one needs to detect is very small (typically ∼0.1 per cent to ∼0.001 per cent). Although this is within the capabilities of modern detectors, when the light from the star passes through the Earth's atmosphere regions of turbulence cause intensity fluctuations (seen as twinkling by the naked eye), known as scintillation. This scintillation, which induces intensity variations in the range of ∼1.0 per cent to 0.1 per cent, limits the detection capabilities of ground-based telescopes (Brown & Gilliland [1994](#); Heasley et al. [1996](#); Ryan & Sandler [1998](#); Charbonneau et al. [2000](#)) to the transits of large exoplanets and prohibits the study of smaller transiting objects.

Recent advances in astronomical adaptive optics (AO) have led to substantial improvements in astronomical imaging; however, there is currently no such correction technique for photometry on large telescopes. The problem is amplified by the fact that the scintillation is predominately caused by high-altitude turbulence so that the range of angles over which the scintillation is correlated is small and normalization by nearby guide stars is impossible. In addition, the intensity of the targets that one wishes to examine is inherently variable and so the correction technique must be independent of any measurement of the object itself.

Here we propose a new method. To correct for the scintillation noise, we intend to make a model of the turbulent atmosphere using computed tomographic algorithms and several guide stars (natural and laser) distributed near to the astronomical target(s). With this model it will be possible to numerically estimate the scintillation signal independently of the target intensity variations and use it to normalize the photometric data. As this technique uses tomographic algorithms to computationally reconstruct the scintillation signal, we can use it to correct the scintillation noise in any direction in the field of view and at any wavelength. We will also be able to correct all objects in the field simultaneously.

As this technique requires several wavefront sensors (WFSs) to probe the turbulent atmosphere, it is closely related to the field of AO tomographic correction for astronomical imaging and spectroscopy. In fact, it will be possible to perform the scintillation correction on any tomographic AO system, either open or closed loop. However, the application of tomography is optimized on larger telescopes where the light cones of the guide stars have a greater overlapping area at altitude and hence have a better sampling of the atmosphere. On a smaller telescope, the guide stars must be close together in order to sample the full atmosphere, but then the angle over which the wavefront aberrations are significantly reduced by the AO system is small. Recent AO upgrades to 8 m class very large telescopes and most AO systems for the next generation of 20–40 m class extremely large telescopes (ELTs) all require tomographic AO systems to function and will therefore make the perfect bases for the technique, requiring no additional hardware.

This technique has the additional benefit that when used in conjunction with an AO system, the image of the field will be corrected. For isolated targets, this increase in Strehl ratio means that there will be a greater contrast between the object and the background, and smaller apertures can be used in the aperture photometry also culminating in a reduction of the overall noise.

[*] E-mail: james.osborn@durham.ac.uk

In crowded fields the situation is more complicated. There is the advantage that the objects will be angularly smaller reducing the field confusion (Esslinger & Edmunds 1998). However, as no AO system is perfect, the residual phase aberration can result in a complicated spatially and temporally varying point spread function (PSF; for example Cagigal & Canales 2000; Currie et al. 2000; Osborn, Myers & Love 2009; Baena Gallé & Gladysz 2011). Even in medium- to high-order correction regimes, quasi-static speckles due to non-common path errors and uncorrected aberrations can add complexity to the PSF (Fitzgerald & Graham 2006; Soummer et al. 2007; Osborn 2012). These variations in the PSF make separating the light from different objects difficult. However, there is interest in solving these problems with advanced PSF modelling and secondary correction techniques (e.g. Véran et al. 1997; Gendron et al. 2006; Turri et al. 2014).

Therefore, AO with scintillation estimation could be competitive with the precision of space-based measurements and the larger collecting areas enabling fainter targets to be observed with higher time resolution photometry. In addition, it is worth noting that this technique will function without the AO system engaged, i.e. only using the AO WFSs and not activating the deformable mirrors (DMs), to allow scintillation correction on non-AO corrected images.

Another technique recently proposed to reduce scintillation noise is conjugate plane photometry (CPP; Osborn et al. 2011). CPP uses a combination of pupil conjugation to the altitude of the dominant scintillation-producing turbulent layer and apodization to remove the scintillation effects. However, CPP is expected to function optimally for telescopes around 2.5 m and not so well for larger telescopes. Here, we propose a solution designed specifically to operate on large and extremely large telescopes.

Section 2 explains the scintillation phenomenon and describes the theory to estimate the noise source for photometric measurements. Section 3 explores the expected magnitude of scintillation noise on photometry. In Section 4 we describe tomography as applied to astronomical AO. In Section 5 we describe the concept, and show results in Section 6.

## 2 THEORY

### 2.1 Scintillation

Optical turbulence results from the mechanical mixing of layers of air with different temperatures and hence density. The refractive index of air depends on its density and so the turbulence creates a continuous screen of spatially and temporally varying refractive indices.

The wavefront from an astronomical source can be considered planar at the top of the Earth's atmosphere. As it propagates to the ground, it becomes aberrated by the atmospheric optical turbulence which forms a limit to the precision of measurements from ground-based telescopes.

The effect of this optical turbulence is twofold. The first effect is to deform the wavefront by retarding the sections passing through regions of higher refractive index. This limits the angular resolution of ground-based telescopes and is a first-order effect as it depends on the first derivative of the wavefront. The second effect of the turbulence is to locally focus and de-focus the wavefront resulting in spatial intensity fluctuations, or speckles, in the pupil plane of a telescope. This is known as scintillation. Fig. 1 shows a simulated pupil image and an actual pupil image from the 2.5 m Isaac Newton Telescope (INT), La Palma. The real image was taken in the visible with an exposure time of 2 ms to freeze the turbulence. Scintillation depends on the curvature induced by the phase screen as it relies on the wavefront being focused and de-focused. Scintillation is a second-order effect as it is dependent on the second derivative of the wavefront. As scintillation is a propagation effect, high-altitude turbulent layers are primarily responsible for the intensity fluctuations. However, phase distortions add linearly and so can have significant components close to the telescope in the boundary turbulent layer.

It is for this reason that a seeing monitor cannot be used to estimate the scintillation noise. In the situation that the surface layer of optical turbulence dominates the seeing aberrations, as is often the case (Osborn et al. 2010), it is entirely possible to have better-than-average seeing and yet worse-than-average photometric conditions. This can happen when the surface layer is weaker than normal or when the high-altitude turbulence is stronger than normal.

These intensity speckles traverse across the pupil with a velocity determined by the velocity of the turbulent layer. Speckles from different layers move independently and superimpose in the pupil plane. As the regions of higher intensity enter and exit the pupil, the integrated intensity also varies. It is these variations which lead to the scintillation noise which can limit the precision of photometric measurements.

The strength of the scintillation is expressed in terms of the scintillation index, $\sigma_I^2$, or the normalized variance of the intensity, assuming no other noise sources,

$$\sigma_I^2 = \frac{\langle I^2 \rangle - \langle I \rangle^2}{\langle I \rangle^2}, \tag{1}$$

where $I$ is the intensity as a function of time and $\langle \rangle$ denotes an ensemble average.

We can derive the theoretical scintillation index as the integral of the scintillation power spectrum, $W(f)$ (Roddier 1981),

$$\sigma_I^2 = \int_0^\infty W(f) \mathrm{d}f, \tag{2}$$

where $W(f)$ is given by (Kornilov 2012)

$$W(f) = 9.7 \times 10^{-3} \times 4 \times (2\pi)^3$$
$$\int_0^\infty C_n^2(z)\phi(f)S(z, f)A(f)f\mathrm{d}z. \tag{3}$$

$S(z, f)$ is the Fresnel filter function to account for the wavefront propagation and is given by $\sin^2(\pi\lambda z f^2)/\lambda^2$ (Roddier 1981), with $\lambda$ being the wavelength of the light. It is this function that gives the intensity fluctuations an intrinsic spatial scale of $r_{\rm F} = \sqrt{\lambda z}$. $A(f)$ is the aperture filter function and is defined by $(2J_1(\pi D f)/(\pi D f))^2$ for a circular aperture (Kornilov 2012). $z$ is the propagation distance to the turbulent layer and the altitude of the layer is then $h = z\cos(\gamma)$, with $\gamma$ being the zenith angle of the observation. $\phi$ is the frequency component of the refractive index power spectrum, for example, for Kolmogorov turbulence $\phi = f^{-11/3}$.

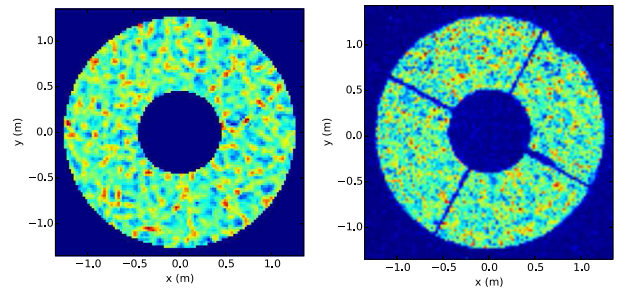

**Figure 1.** Example simulated pupil image for a 2.5 m telescope and a single turbulent layer at 10 km (left) and an example of a real pupil from the INT (2.5 m), La Palma (right). The observed image was recorded in the visible with an exposure time of 2 ms.

The scintillation index can be calculated from atmospheric and telescope parameters as (Dravins et al. 1998)

$$\sigma_{I,\rm se}^2 = 17.34 D^{-7/3}(\cos\gamma)^{-3}\int_0^\infty h^2 C_n^2(h)\,{\rm d}h, \qquad (4)$$

for short exposures, and

$$\sigma_{I,\rm le}^2 = 10.66 D^{-4/3} t^{-1}(\cos\gamma)^\alpha \int_0^\infty \frac{h^2 C_n^2(h)}{V_\perp(h)}\,{\rm d}h, \qquad (5)$$

for long exposures, where $D$ is the telescope diameter, $t$ is the observation exposure time, $C_n^2(h)$ is the refractive index structure constant and is a measure of the turbulence strength, $V_\perp(h)$ is the wind velocity profile and $\alpha$ is the exponent of the airmass. Note that the value of the airmass exponent, $\alpha$, will depend on the wind direction and vary between $-3$ for the case when the wind is transverse to the azimuthal angle of the star and $-4$ in the case of a longitudinal wind direction. This is a geometric correction. In the case where the wind direction is parallel to the azimuthal angle of the star, the projected pupil on to a horizontal layer is stretched by a factor of $1/\cos\gamma$; this changes the projected wind speed.

The characteristic correlation scale of the intensity fluctuations at the ground is given by the radius of the first Fresnel zone, $r_{\rm F} = \sqrt{z\lambda}$. As a wavefront propagates away from a turbulent layer, increasing $z$, the spatial intensity fluctuations become larger both in terms of intensity and spatial extent, increasing $r_{\rm F}$. This is not dependent on the strength of the layer which only affects the magnitude of the intensity fluctuations and not their spatial properties.

For small telescopes, where the aperture size is smaller than the spatial scale of the intensity fluctuations ($D < r_{\rm F}$), there is not enough spatial averaging to remove the dependence on wavelength. It is only when many speckles are spatially averaged within the pupil that the small-angle approximation can be invoked on the Fresnel filter function ($\sin^2(\pi\lambda h f^2)/\lambda^2 = (\pi\lambda h f^2)^2/\lambda^2$). The scintillation index for small telescopes can be approximated by (Dravins et al. 1998)

$$\sigma_I^2 = 19.2\lambda^{-7/6}(\cos\gamma)^{-11/6}\int_0^\infty h^{5/6}C_n^2(h)\,{\rm d}h. \qquad (6)$$

This regime is not developed further as we are primarily interested in correcting scintillation on large telescopes. It is only shown here for completeness.

## 3 SCINTILLATION IN PHOTOMETRY

To estimate the effect of the scintillation on photometric measurements, we can use equations (4) and (5) with an estimate of the atmospheric turbulence profile at a given site. Scintillation correction is only required in scenarios where the scintillation noise is significant. Here we take that shot noise to be the fundamental limit to the precision and assume other noise sources to be negligible.

Using recent results from a new scintillation detection and ranging (SCIDAR) atmospheric optical turbulence profiling instrument, Stereo-SCIDAR (Osborn et al. 2013; Shepherd et al. 2013), over 30 nights distributed between 2013 May and September, on La Palma, we can calculate the median atmospheric turbulence profile and then use this to estimate the effective strength at a single turbulent layer at 10 km in order to conserve the median scintillation index. This is done to simplify the calculations and simulations for the rest of this work.

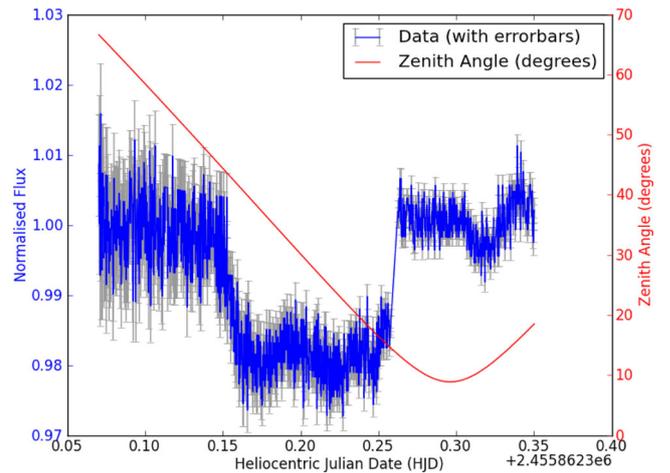

**Figure 2.** WASP33 transit taken on the 0.5 m telescope on La Palma. Also shown is the concurrent zenith angle. The data preceding the transit are more noisy due to the airmass dependence of scintillation.

The measured atmospheric optical turbulence profiles show a median free atmosphere turbulence strength of $C_n^2(h){\rm d}h = 105 \times 10^{-15}$ m$^{1/3}$ with a first and third quartile of $64 \times 10^{-15}$ m$^{1/3}$ and $174 \times 10^{-15}$ m$^{1/3}$, respectively. Using equation (5) we can estimate the scintillation noise on a large telescope. If we assume an 8 m telescope and a zenith distance of 30 deg and constrain the median turbulence profile from Stereo-SCIDAR to 10 km and an exposure time of 10 s, then the scintillation rms noise will be approximately 0.03 per cent. If we assume a target of magnitude 10, then the shot noise of the observations would be 0.006 per cent. Therefore, it can be seen that there is significant gains that can be exploited if we could reduce the scintillation noise. For example, with rms noise levels less than 0.01 per cent it would be possible to observe the transit of Earth-sized extrasolar planets. In this case, the shot noise will be equal to the scintillation noise when observing a magnitude 13.5 target. Therefore, the technique will reduce the total photometric noise for all targets brighter than magnitude 13.5. This limiting magnitude will vary depending on atmospheric conditions, telescope parameters and zenith angle.

In addition, the scintillation noise dependence on the zenith distance is important as it means that the scintillation noise will change during an observation. It will be higher for a larger airmass. Fig. 2 shows an example transit of WASP33 measured on a 0.5 m telescope on La Palma. We see a correlation between the noise and the airmass, as expected. This is especially important for observations which require stable photometry for long durations, for example transits of extrasolar planets where one wishes to measure the ingress and egress of the planetary transit as well as a substantial period of time on either side in order to constrain model fits to the data.

## 4 AO TOMOGRAPAHY

AO systems require reference sources to sample the turbulent atmosphere above the telescope. If a guide star is located very close to the target or we can use the target itself, then this star can be used to directly measure the phase aberrations along the line of sight to the target. However, if there is no guide star bright enough or we would like to observe multiple or extended objects in the field, then we require multiple guide stars to sample the volume of turbulence above the telescope. If the light cones of these guide stars overlap with the cylinder projected to the target, we can use tomographic techniques to reconstruct the phase aberrations along the line of sight to the target. The majority of modern AO systems (with the exception of extreme AO for extrasolar planet imaging) make use of tomographic reconstruction techniques. Three major varieties of tomographic reconstruction currently under investigation are laser tomography AO (Le Louarn & Hubin 2004), multiconjugate AO (MCAO; Beckers 1989) and multi-object AO (MOAO; Hammer et al. 2002; Assémat et al. 2007; Morris et al. 2013).

In AO, tomographic reconstruction is the re-combination of the information from several guide stars to estimate the phase aberrations along a different line of sight to a scientific target. A standard approach is to use WFSs to measure the summed phase aberrations along the line of sight to the guide stars. Where the light cones overlap at the altitude of a turbulent layer, the same phase aberrations will be applied to both wavefronts but in different areas of the meta-pupil. We can then look for correlation in the phase maps at the ground. Fig. 3 shows a topological diagram of a system with three guide stars and one target. Any turbulence at low altitudes will be well sampled. At higher altitudes the overlap is reduced and we therefore have less information. Above the altitude where the beams no longer overlap, there will be very limited correlation in the phase aberration (possibly some correlation in the very low order modes, depending on the extent of the separation and the outer scale of the turbulence), and it is therefore difficult to gain any information. Any turbulence above this altitude will essentially add noise to the measurements. This also explains why tomography works better on larger telescopes. Smaller telescopes would require the guide stars to be close together in order to be overlapping through the entire turbulent atmosphere and this will significantly reduce the corrected field of view of the wide-field AO system.

AO tomography can be used to estimate low-order phase aberrations in the turbulent layers. The standard least-squares method of tomographic reconstruction (e.g. Ellerbroek 1994; Fusco et al. 2001) involves multiplying the WFS vectors with a control matrix. The control matrix maps the response of the WFSs to the actuator commands of the DMs and can be computed off-sky. In the case of MCAO, several DMs are placed at some conjugate altitude in the optical light path and are used to correct the turbulence at that altitude. In MOAO we can place 'virtual' DMs at the conjugate altitude of the turbulent layers in the atmosphere and calculate the control matrix of each of these independently by using either telescope simulator on an optical bench or via simulation. It is very important to position these virtual DMs accurately at the conjugate altitude of the turbulent layers or the performance will be compromised (Gendron et al. 2014). If the profile of optical turbulence were to change during observation, the tomographic reconstructor would provide a poor fit to the actual slopes. We therefore require high vertical resolution atmospheric optical turbulence profiles in order to optimize this tomographic reconstructor.

We can estimate the atmospheric optical turbulence profile using the AO telemetry from the WFSs using the slope detection and ranging (SLODAR) method with either natural guide stars (Wilson 2002) or the laser guide stars (LGS; Cortés et al. 2012). SLODAR works by correlating the wavefront slopes from two or more WFSs, and a turbulent layer will appear in the covariance function as a peak with an offset from the centre proportional to the altitude of the turbulent layer and a magnitude proportional to the strength of the layer. Alternatively, an external atmospheric profiler could be used to measure the turbulence profile.

There have been several developments recently towards AO tomographic reconstructors which are adaptive or insensitive to changing atmospheric conditions (for example, Vidal, Gendron & Rousset 2010; Osborn et al. 2014). However, here we assume an optimal tomographic reconstructor (i.e. one that is optimized for the current atmospheric conditions), and the effect under investigation is the fitting error of the wavefront due to the finite sampling of the wavefronts.

## 5 SCINTILLATION REDUCTION

Scintillation noise is caused by intensity fluctuations over several decades of scale. Fig. 4 shows a normalized theoretical power spectrum for the scintillation, as calculated using equation (3). This assumes a circular aperture without a secondary obscuration. The parameters are chosen so that the cutoff frequency and the magnitude of the power spectrum are equal to unity. We emphasis here that the scintillation noise does have power at low frequencies and is not restricted to high-frequency perturbations.

The low-order fluctuations are caused by low-order curvature in the turbulent layer. A qualitative explanation of this is to consider a region of turbulence through which the light passes. If this area of turbulence has a net focus aberration, then there will be an increase in the photon density and hence more photons collected by the telescope, and vice versa. Fig. 5 shows a schematic diagram of this concept.

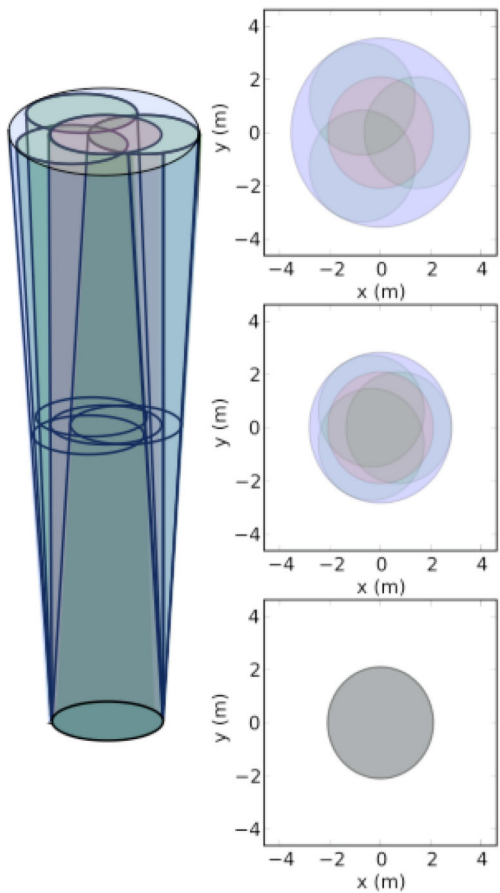

**Figure 3.** Topological diagram of the light cones for three guide stars and one target for a 4.2 m telescope and guide stars equally distributed on a ring of radius 30 arcsec. The target direction is shown in red, the guide stars in green and the full field of view in blue. The cut-throughs on the right are taken at 0, 5000 and 10 000 m. At higher altitudes the overlap of the guide stars reduces and we sample smaller areas of the target light cone.

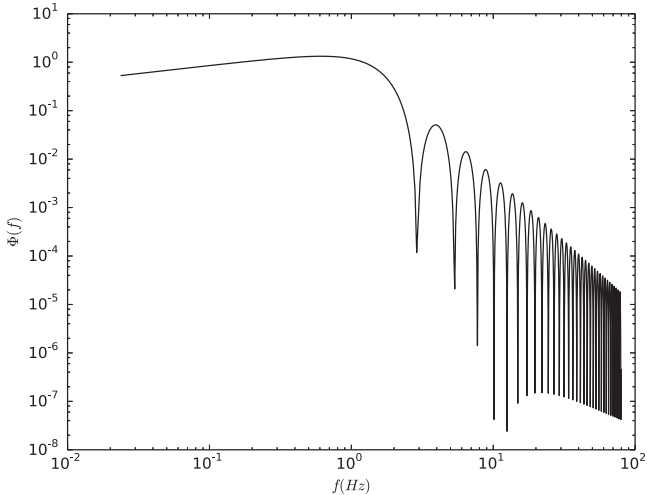

**Figure 4.** Theoretical temporal power spectrum of scintillation. Scintillation shows power on all temporal scales. In a real system, the oscillations at high temporal frequency, $f$, will be smeared out by wind velocity variations (gusting) and propagation through multiple turbulent layers.

Once the tomographic reconstructor has been optimized for the concurrent atmospheric conditions, it can be used to estimate the wavefront aberration at the altitude of the turbulence. For MCAO systems, the DM shapes can be used as the atmospheric model. For MOAO systems, the atmospheric aberrations must be reconstructed via, for example, the virtual DM method. In order to realize the scintillation correction, this wavefront can then be numerically propagated to the ground via all of the detected and reconstructed layers.

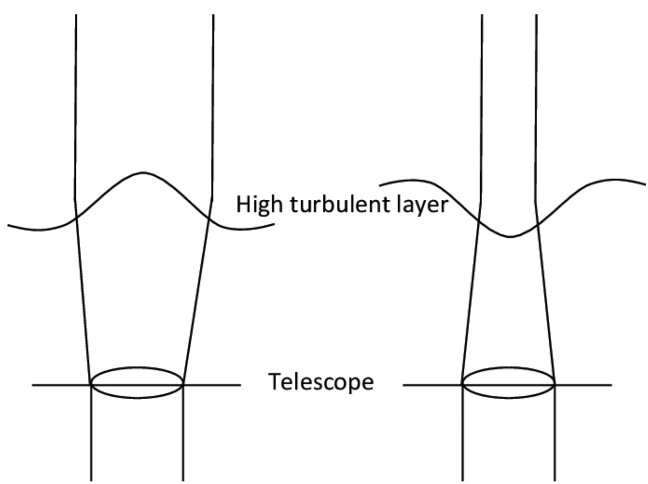

**Figure 5.** Schematic diagram to show the effect of low-order phase aberrations on the footprint of the beam. Focus in a high-altitude layer will concentrate the speckles into a smaller area, effectively increasing the amount of light seen by the telescope. De-focus will spread the beam reducing the amount of light collected by the telescope.

For a single layer the wavefunction, $\Psi$, at the telescope pupil is given by

$$\Psi(x, y) = \left[ K(z) \otimes \exp(\mathrm{i}\phi_h) \right] P(x, y), \tag{7}$$

where $z$ is the propagation distance, $x$ and $y$ are spatial coordinates, $P(x, y)$ is the telescope pupil function, $\phi_h$ is the turbulent phase function at altitude $h$ which is illuminated by the cone of light between the telescope and the target, $\otimes$ denotes a convolution and $K$ is the Fresnel propagation kernel, given by

$$K(z) = \frac{\mathrm{i}}{\lambda z} \exp(\mathrm{i}kz) \exp\left( \frac{\mathrm{i}k}{2z} \left[ \left(x - x'\right)^2 + \left(y - y'\right)^2 \right] \right), \tag{8}$$

where $k$ is the wavenumber, $\lambda$ is the wavelength of the light and $x'$ and $y'$ are spatial coordinates in the observation plane located at a distance $z$. Positive $z$ indicates a diverging spherical wavefront and negative $z$ is a converging spherical wavefront or a positive and negative propagation, respectively. For multiple layers, the wavefunction is propagated from the highest layer to the ground via each intermediate layer,

$$\Psi(x, y) = K(z_2) \otimes \left[ \left[ K(z_1) \otimes \exp(\mathrm{i}\phi_{h_1}) \right] \exp(\mathrm{i}\phi_{h_2}) \right] P(x, y), \tag{9}$$

where $z_1$ and $z_2$ are the propagation distances from turbulent layers at altitudes $h_1$ to $h_2$ and $h_2$ to the ground, respectively. The spatially and temporally integrated intensity over the pupil and the exposure time can then be used to normalize the photometric light curve of the observation.

To estimate the expected scintillation correction, we assume that the AO system acts as a high-pass spatial filter removing phase aberrations with a spatial frequency below $f_{\mathrm{AOcutoff}} \approx 1/d$. Therefore, we can use the AO system to estimate the low-frequency aberrations and calculate the associated wavefront curvature. The temporal cutoff frequency is then $f_{\mathrm{cutoff}} \approx V_\perp/d$, and low-order intensity variations in the light curve below this frequency are removed. The uncorrected temporal power spectrum will have a cutoff at $f_{\mathrm{cutoff}} \approx V_\perp/D$ (Fig. 4).

## 6 RESULTS

For the purpose of this study, we have chosen to simulate an 8 m telescope. Five LGS WFSs are used to probe the atmosphere. The LGS asterism has an LGS at each corner of a square with off-axis angle of 45 arcsec and one LGS in the centre. This asterism was chosen to ensure that light cones overlap up to 15 km, which means that all significant turbulence is sampled. A single on-axis natural guide star is used for tip-tilt correction although this information is not used for the scintillation correction. We assume perfect knowledge of the structure of the turbulence, i.e. that it has a Von Karman power spectrum with an outer scale of 30 m.

The atmosphere has two turbulent layers, one at the ground and one at altitude (at 10 km, unless stated otherwise). Each layer has an effective $r_0 = 0.32$ m $[C_n^2(h)\mathrm{d}h = 1.05 \times 10^{-13}$ m$^{1/3}$, which is equal to the free atmosphere median of La Palma as measured by Stereo-SCIDAR (Shepherd et al. 2013)], culminating in an integrated value of $r_0 = 0.20$ m (0.5 arcsec seeing) and a wind speed of 15 m s$^{-1}$. However, as scintillation is a second-order effect, it requires the propagation of the wavefront and so the ground layer results in no scintillation noise. We assume that observations are at zenith.

Fig. 6 shows an example simulated pupil irradiance pattern from a turbulent layer at 10 km and an 8 m telescope. Although it appears that this pupil pattern is dominated by intensity fluctuations of the scale $r_F$, we have seen from previous sections that there are also low-order intensity variations. Fig. 7 shows an example simulated pupil image of the reconstructed wavefront at 10 km which has been numerically propagated to the ground. As the reconstructed wavefront only has low-order phase aberrations, the intensity pattern at the ground also appears to be filtered, with any high-order spatial fluctuations removed. We then integrate the intensity in the reconstructed pupil image over the exposure time of the science camera which results in an estimate of the scintillation magnitude for that observation. This can be used to correct for the scintillation signal, irrespective of any change in intensity of the source.

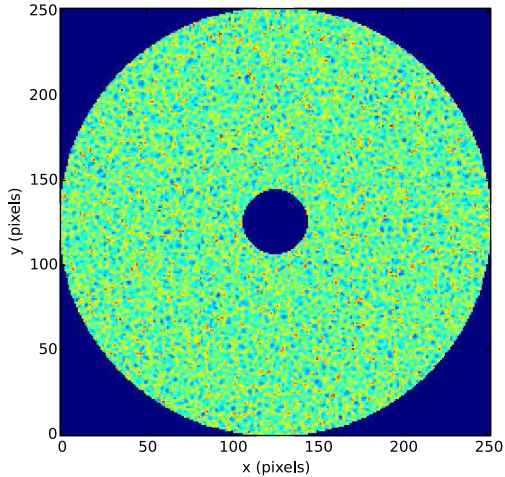

**Figure 6.** Example simulated pupil image for an 8 m telescope and a turbulent layer at 10 km.

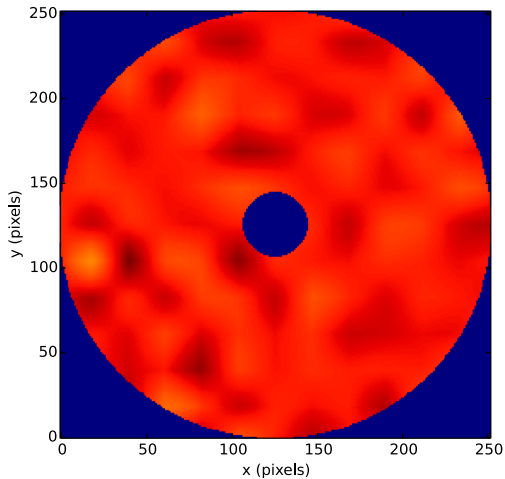

**Figure 7.** Example simulated scintillation correction pupil image for an 8 m telescope and a turbulent layer at 10 km.

Fig. 8 shows an example light curve with and without scintillation correction from a simulation of a 16 × 16 system for 5 s of simulation time. The high-altitude wavefront is also reconstructed into 16 × 16 sub-apertures. We see that the low-frequency intensity fluctuations have been removed and only high-frequency components remain. These high-frequency fluctuations cannot be sensed with the 16 × 16 AO system. A higher order AO system would detect higher order intensity variations and lead to better scintillation correction. In this case, the scintillation noise has been reduced from 0.2 to 0.05 per cent.

Fig. 9 shows a simulated light curve for ~80 min of simulated time, consisting of 500 exposures of 10 s of simulated exposure time. The scintillation rms noise has been reduced by over a factor of 10.

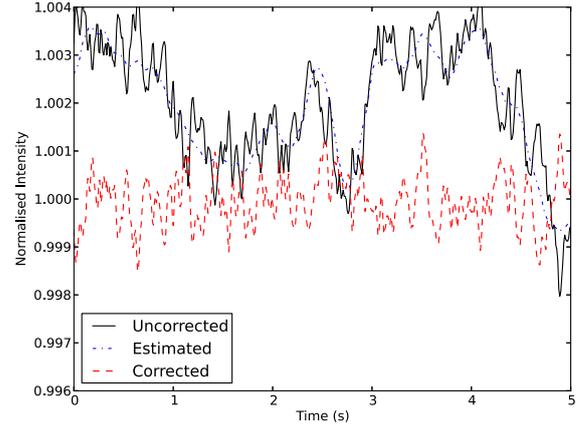

**Figure 8.** Example light curve for an 8 m telescope, 16× 16 sub-apertures, 0.02 s simulated exposure time, 5 s total simulation time and a turbulent layer of strength $1.05 \times 10^{-13}$ m$^{1/3}$ at 10 km. The solid line shows the measured light curve, the dot–dashed line shows the estimated light curve from the reconstructed wavefront and the dashed line indicates the corrected light curve. The scintillation noise has been reduced from 0.2 to 0.05 per cent.

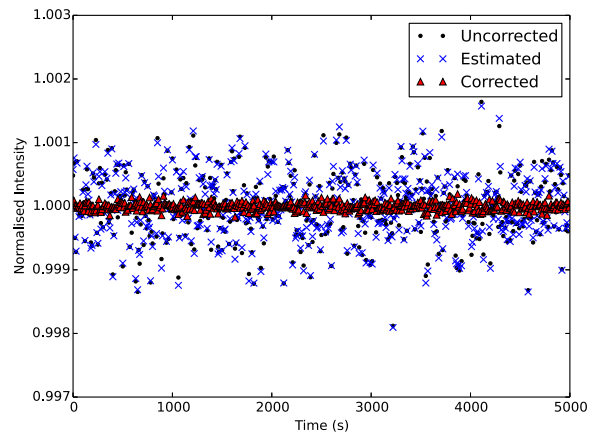

**Figure 9.** Simulated long-duration light curve for a 16 × 16 AO system with five guide stars, four 45 arcsec off-axis and one in the centre on an 8 m telescope. The scintillation noise has been reduced from 0.04 to 0.003 per cent for 80 min of data and 10 s exposure time. The dots are the measured intensity integrated over 10 s, the crosses are the estimated scintillation signal and the triangles denote the corrected light curve.

The temporal power spectrum of the scintillation intensity variations (Fig. 10) shows that the corrected power spectrum converges at $f_{\rm knee}$. This frequency, which depends on the AO system parameters, determines the effectiveness of the technique. If perfect knowledge of the phase aberration could be determined, $f_{\rm knee} \rightarrow \infty$ and $\sigma_{\rm scint} \rightarrow 0$.

As the scintillation noise is proportional to the integrated power spectrum, we expect a better improvement ratio for longer exposure times and for AO systems with lower tomographic error. From the power spectrum we see that the uncorrected curve has a cutoff at ~$V_\perp/D$; in this case $f_{\rm knee} \sim 2$ s$^{-1}$. The scintillation corrected power spectra converge at $f_{\rm knee} \approx V_\perp/d \sim 15$ and 30 s$^{-1}$ for the 8 × 8 and 16 × 16 correction, respectively.

In the Monte Carlo simulation, the scintillation rms noise has been reduced from 0.04 per cent to 0.006 and 0.003 per cent for 8 × 8 and 16 × 16 correction, respectively. However, reducing the scintillation noise is only beneficial if it makes a significant contribution to the noise budget. In the case shown the shot noise from the target will be equal to the uncorrected scintillation noise when observing a target of stellar magnitude ∼14.5. Therefore, neglecting all other noise sources, the technique will reduce the total photometric noise for targets brighter than this magnitude. This also assumes the median free atmospheric conditions on La Palma as measured by one campaign. The technique will be beneficial for fainter target magnitudes when the scintillation noise is higher. This will occur on occasions when the mid to high-altitude turbulence is stronger or when observing at larger zenith angles. We assume other noise sources to be negligible, which means that in reality great care would still have to be taken to reach the performance quoted here.

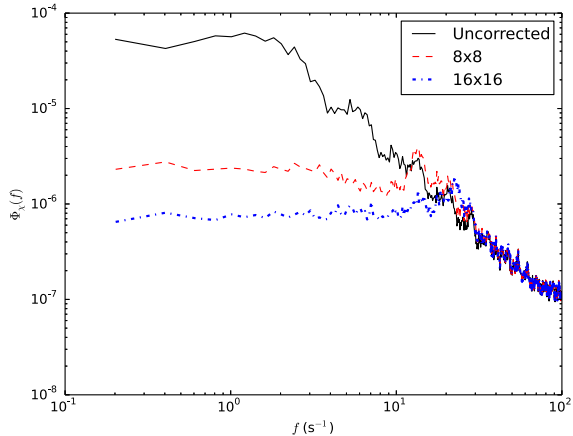

**Figure 10.** Scintillation power spectrum for Kolmogorov turbulence. The solid line denotes the uncorrected scintillation power spectrum, the dashed and dash–dotted lines are the residual power spectrum after $8 \times 8$ and $16 \times 16$ correction, respectively.

## 7 CONCLUSION

Scintillation noise caused by the Earth's atmosphere can limit the precision of ground-based astronomical photometry. By building a model of the atmospheric turbulence above the telescope using a tomographic algorithm, we can estimate the scintillation noise in any target direction and at any wavelength. This estimate is used to calibrate the photometric light curve and reduce the noise. This system can be used with any tomographic wide-field AO system, including all modern systems being implemented on the current generation of very large (∼8 m) and the next generation of ELTs, using natural guide stars or LGS to probe the turbulent atmosphere. The exact improvement ratio that can be obtained using this method of tomographic correction will depend on the geometry of the AO system being used; however, for large telescopes we expect the photometric noise to be reduced by an order of magnitude. In simulation we have seen that the scintillation noise is reduced by a factor of 5 for an $8 \times 8$ AO system and 10 for a $16 \times 16$ system on an 8 m telescope with four LGS on a ring of radius 45 arcsec and one in the centre of the field. Therefore, AO with scintillation estimation could be competitive with the precision of space-based measurements and the larger collecting areas enabling fainter targets to be observed with higher time resolution photometry.

## ACKNOWLEDGEMENTS

I would like to thank Richard Wilson for very helpful comments during the preparation of this paper and also Tim Butterley and the pt5m team for the transit data. This work was supported by the Science and Technology Facilities Committee (STFC) [ST/J001236/1]. The Isaac Newton Telescope is operated on the island of La Palma by the Isaac Newton Group in the Spanish Observatorio del Roque de los Muchachos of the Instituto de Astrofísica de Canarias.

## REFERENCES

Assémat F., Gendron E., Hammer F., 2007, MNRAS, 376, 287
Baena Gallé R. B., Gladysz S., 2011, PASP, 123, 865
Beckers J. M., 1989, Proc. SPIE, 1114, 215
Brown T. M., Gilliland R. L., 1994, ARA&A, 32, 37
Cagigal M. P., Canales V. F., 2000, J. Opt. Soc. Am. A, 17, 1312
Charbonneau D., Brown T. M., Latham D. W., Mayor M., 2000, ApJ, 529, L45
Cortés A., Neichel B., Guesalaga A., Osborn J., Rigaut F., Guzman D., 2012, MNRAS, 427, 2089
Currie D. et al., 2000, in Manset N., Veillet C., Crabtree D., eds, ASP Conf. Ser. Vol. 216, Astronomical Data Analysis Software and Systems IX. Astron. Soc. Pac., San Francisco, p. 381
Dravins D., Lindegren L., Mezey E., Young A. T., 1998, PASP, 110, 610
Ellerbroek B. L., 1994, J. Opt. Soc. Am. A, 11, 783
Esslinger O., Edmunds M. G., 1998, A&A, 129, 617
Fitzgerald M. P., Graham J. R., 2006, ApJ, 637, 541
Fusco T., Conan J., Rousset G., Mugnier L. M., Michau V., 2001, J. Opt. Soc. Am. A, 18, 2527
Gendron E., Clénet Y., Fusco T., Rousset G., 2006, A&A, 457, 359
Gendron E., Morel C., Osborn J., Martin O., Gratadour D., Vidal F., Le Louarn M., Rousset G., 2014, Proc. SPIE, 9148, 91484N
Hammer F. et al., 2002, in Bergeron J., Monnet G., eds, Scientific Drivers for ESO Future VLT/VLTI Instrumentation. Springer-Verlag, Berlin, p. 139
Heasley J. N., Janes K., LaBonte B., Guenther D., Mickey D., Demarque P., 1996, PASP, 108, 385
Kornilov V., 2012, MNRAS, 426, 647
Le Louarn M., Hubin N., 2004, MNRAS, 349, 1009
Morris T. et al., 2013, in Esposito S., Fini L., eds, Proc. Third AO4ELT Conference, Multiple Object Adaptive Optics: Mixed NGS/LGS Tomography. Available at http://ao4elt3.sciencesconf.org
Osborn J., 2012, MNRAS, 424, 2284
Osborn J., Myers R. M., Love G. D., 2009, Opt. Express, 17, 17279
Osborn J., Wilson R. W., Butterley T., Shepherd H., Sarazin M., 2010, MNRAS, 406, 1405
Osborn J., Wilson R. W., Dhillon V., Avila R., Love G. D., 2011, MNRAS, 411, 1223
Osborn J., Wilson R. W., Shepherd H., Butterley T., Dhillon V. S., Avila R., 2013, in Esposito S., Fini L., eds, Proc. 3rd AO4ELT Conference, Stereo SCIDAR: Profiling Atmospheric Optical Turbulence with Improved Altitude Resolution. Available at http://ao4elt3.sciencesconf.org/
Osborn J. et al., 2014, MNRAS, 441, 2508
Roddier F., 1981, in Wolf E., ed., Progress in Optics, Vol. 19: The Effects of Atmospheric Turbulence in Optical Astronomy. North-Holland, Amsterdam, p. 281
Ryan P., Sandler D., 1998, PASP, 110, 1235
Shepherd H., Osborn J., Wilson R. W., Butterley T., Avila R., Dhillon V., Morris T. J., 2013, MNRAS, 437, 3568
Soummer R., Ferrari A., Aime C., Jolissaint L., 2007, ApJ, 669
Turri P., McConnachie A. W., Stetson P. B., Fiorentino G., Andersen D. R., Bono G., Véran J.-P., 2014, in Marchetti E., Close L. M., Véran J.-P., eds, Proc. SPIE Vol. 9148, Photometric Performance of LGS MCAO with Science-Based Metrics: First Results from Gemini/GeMS Observations of Galactic Globular Clusters. SPIE, Bellingham, p. 91483V
Véran J.-P., Rigaut F., Maitre H., Rouan D., 1997, J. Opt. Soc. Am. A, 14, 3057
Vidal F., Gendron E., Rousset G., 2010, J. Opt. Soc. Am. A, 27, 253
Wilson R. W., 2002, MNRAS, 337, 103

This paper has been typeset from a TeX/LaTeX file prepared by the author.